\documentclass[10pt,letterpaper]{IEEEtran}
\usepackage{stmaryrd}
\usepackage{color}
\usepackage{url}
\usepackage[latin9]{inputenc}
\usepackage{amsthm}
\usepackage{amsmath}
\usepackage{amsthm}
\usepackage{amssymb}
\usepackage{graphicx}
\usepackage{epstopdf}
\usepackage{subcaption}
\usepackage{wrapfig}
\usepackage{algorithm}
\usepackage{algpseudocode}
\usepackage{mathrsfs}
\usepackage{float}
\usepackage{mathtools}
\usepackage{tabulary}
\usepackage{booktabs}
\usepackage{caption}
\usepackage{xprintlen}
\usepackage{multirow}
\usepackage{color}
\usepackage{bm}

\newcommand{\algo}{{\texttt {HiNoVa}}}

\newcommand{\aalgo}{{\texttt {HiNoVa}}}

\PassOptionsToPackage{bookmarks={false}}{hyperref}

\IEEEoverridecommandlockouts

\def\BibTeX{{\rm B\kern-.05em{\sc i\kern-.025em b}\kern-.08em
   T\kern-.1667em\lower.7ex\hbox{E}\kern-.125emX}}

\newcommand{\comment}[1]{ }

\newcommand\subparagraph{%
  \@startsection{subparagraph}{0}
  {\parindent}
  {0ex \@plus 0ex \@minus 0ex}
  {-1em}
  {\normalfont\normalsize\bfseries}}
\makeatother

\usepackage{titlesec}
\titlespacing*{\section}
{0pt}{1.2ex}{0ex} 
\titlespacing*{\subsection}
{0pt}{1ex}{0ex}  
\titlespacing*{\subsubsection}
{0pt}{0ex}{0ex} 

\begin{document}

\title{\aalgo: A Novel Open-Set Detection Method for Automating RF Device Authentication}


\author{Luke Puppo, Weng-Keen Wong, Bechir Hamdaoui, Abdurrahman Elmaghbub ~\\
 School of EECS, Oregon State University, Corvallis, OR, USA ~\\ 
 \{puppol, wongwe, hamdaoui, elmaghba\}@oregonstate.edu
 
}

\maketitle
\thispagestyle{plain}
\pagestyle{plain}

\begin{abstract}
New capabilities in wireless network security have been enabled by deep learning, which leverages patterns in radio frequency (RF) data to identify and authenticate devices. 
Open-set detection is an area of deep learning that identifies samples captured from new devices during deployment that were not part of the training set. Past work in open-set detection has mostly been applied to independent and identically distributed data such as images. In contrast, RF signal data present a unique set of challenges as the data forms a time series with non-linear time dependencies among the samples. We introduce a novel 
open-set detection approach based on the patterns of the hidden state values within a Convolutional Neural Network (CNN) Long Short-Term Memory (LSTM) model. Our approach greatly improves the Area Under the Precision-Recall Curve on LoRa, Wireless-WiFi, and Wired-WiFi datasets, and hence, can be used successfully to monitor and control unauthorized network access of wireless devices.
\end{abstract}



\section{Introduction}
\label{intro}
The proliferation of Internet Of Things (IoT) devices in sensitive environments, such as military bases, government buildings, and private businesses, creates a need for detecting anomalous devices that pose security threats. These devices can easily bypass security measures as they can be concealed. Traditional detection methods are ineffective at identifying unauthorized wireless devices, especially with attacks like cloning and man-in-the-middle \cite{5601960}.

RF fingerprinting is a recognized key method to enhance security in IoT networks~\cite{deployment_variability}. It extracts device-specific features from RF signals to identify wireless transmitters, leveraging unique hardware imperfections during transmitter manufacturing. Feature extraction methods range from hand-crafted to deep learning-based approaches that identify features from raw RF signals. This paper proposes \aalgo, a new machine learning-based open-set detection method that identifies unauthorized (also referred to as unknown or unseen) IoT devices and authorized (also referred to as known or seen) devices. \aalgo~is tested on datasets collected from devices using LoRa and WiFi protocols. LoRa is a wireless communication technology designed for IoT devices that operates in the sub-gigahertz frequency range, enabling long-range, low-power, bi-directional communication. LoRa's advantages include longer range, better penetration through obstacles, and low power consumption, making it suitable for IoT applications that require a wide area network coverage. However, LoRa has lower data rates than WiFi, making it unsuitable for high-speed data transfer applications. The crowded sub-gigahertz frequency range can also lead to interference from other devices. Each of the protocols, LoRa and WiFi, has its practical use and is commonly adopted by various transmitters, and hence, our proposed open-set detection method is tested using both protocols. 

\subsection{Open-Set Detection and Device Authentication}
Supervised machine learning algorithms typically operate under \emph{closed-set} recognition, meaning that they assume the classes encountered during testing are identical to those seen during training. This means that if a Neural Network (NN) is trained to identify the two classes of cats and dogs, it fails to recognize an unknown type of animal, such as a bird, as a distinct animal and will instead misclassify it as either a cat or a dog. This limitation is particularly problematic in real-world scenarios where wireless device fingerprinting is used for security purposes. In this security use case, the classes correspond to \emph{known} devices and it is crucial for the system to accurately detect \emph{unknown} devices (i.e. the open-set devices) to raise an alert.
For this type of problems, \emph{Open-set} detection \cite{Scheirer2013} can be used, where the classifier needs to recognize that data samples do not belong to any of the known devices seen during training, and raises an alert when this happens. Our work introduces \algo, a novel open-set detection approach for authenticating wireless devices using RF fingerprinting.


\subsection{Related Work}

One of the simplest approaches to open-set detection is to use the predicted class probability as an indicator of the model's confidence that the data instance belongs to one of the known devices \cite{hendrycks_baseline_2017}. In a NN, the predicted class probability is the maximum class probability output by a softmax distribution. If this value is low, it indicates that the instance is likely from an unknown device.

Recent work \cite{vaze22,dietterich22} shows that the maximum logit score (which we refer to as \emph{MaxLogit}) is a stronger baseline for detecting open-set instances. Logits are the outputs of the last linear layer of a deep neural network. In classification, these logits are the inputs to the softmax layer, which normalizes the logits to be a valid probability. Normalizing the logits removes information about their raw magnitude, which is valuable for detecting open-set instances \cite{vaze22}. The MaxLogit score is the value of the largest logit, which is indicative of the uncertainty of the classifier as to the device; an open-set instance should have a lower maximum logit value. 

Recent approaches to open-set detection focus on leveraging internal node values and activation patterns of neurons inside neural networks to detect open-set samples. For example, ReAct \cite{Sun21} analyzes the internal activations of neural networks and identifies highly distinctive signature patterns for open-set distributions. Dietterich et al. \cite{dietterich22} argue that detecting novel objects in object recognition applications with an open set of possible categories is a familiarity-based problem rather than a novelty-based problem. Their familiarity hypothesis posits that state-of-the-art methods based on the computed logits of visual object classifiers succeed by detecting the absence of familiar learned features rather than the presence of novelty. 

Much of the literature for open-set detection applies to data instances that are independent and identically distributed (i.i.d). To our knowledge the only work for open-set detection on time series is by Akar et al. \cite{akar_open_2022}, which clusters the time series in each known class to identify a class-specific barycenter; then, during deployment, new time series are identified by how close they are to these barycenters, where the closeness is determined by dynamic time warping (DTW) and also by cross-correlation. Time series that are not close to the barycenters of known devices are flagged as an unknown device. DTW has a complexity of $O(T^2)$, where $T$ is the length of the two time series to be aligned. The algorithm by Akar et al. uses DTW in the inner loop of several operations and is extremely computationally expensive.

A handful of papers have applied open-set detection to RF fingerprinting. Gritsenko et al. \cite{Gritsenko2019} use the maximum probability from the softmax layer and the ratio of slices predicted to belong to each device to establish the confidence in the device prediction. Hanna et al. \cite{Hanna2020} investigate a variety of methods such as the maximum softmax probability and methods that incorporate data from known unauthorized devices. 
Gaskin et al.~\cite{gaskin2022tweak} proposes Tweak, a lightweight calibration approach that leverages metric learning to achieve high open-set accuracy without the need for model re-training, making it more suitable for resource-constrained applications.
In a recent work, Karunaratne et al.~\cite{Karunaratne2021} use generative deep learning models to produce synthetic data from unauthorized devices, which are used to augment the training set. Our approach differs from these approaches by modeling the time series nature of the data with a CNN+LSTM and performing open-set detection.
Another closely related area to open-set detection is anomaly detection \cite{Chandola2009}. In anomaly detection, the goal is to identify individual outliers that are rare with respect to the "normal" data instances. Anomaly detection has some subtle differences with open-set detection. First, in open-set detection, data instances from the unknown class come from a semantically coherent grouping that is different from the known classes. In contrast, the anomalies found by anomaly detection need not form a coherent grouping. Second, the anomalies in a typical anomaly detection setting make up a small fraction of the data, with the "normal" instances forming a large proportion of the data. In open-set detection, the unknown classes can potentially contain a large number of data instances. Despite these subtleties, anomaly detection techniques can, in some cases, be applied to open-set detection and vice versa; however, open-set detection methods generally outperform anomaly detection methods for detecting unknown devices \cite{hanna_open_2021}.

\subsection{Contributions}
We introduce  \algo, a novel open-set detection method for wireless communication protocols. \algo~leverages the \texttt{Hidden Node Values} within a trained Long-Short-Term Memory (LSTM) unit of  a deep NN to generate a unique device fingerprint for each known device. Then, new fingerprints encountered during deployment can be compared against the fingerprints of known devices, enabling the system to accurately identify unknown devices. After undergoing training on a set of known devices, the open-set detection process is highly efficient and can be performed in real-time even on consumer-grade devices. This makes \algo~an ideal solution for wireless security applications, where the ability to quickly identify unauthorized/unknown devices is of utmost importance.

The paper is structured as follows: Section~\ref{back} presents the machine learning architecture used by our method. Section~\ref{sec:sys} presents the details of the \algo~algorithm. Section~\ref{sec:dataset} describes the LoRa, Wireless-WiFi, and Wired-WiFi datasets used in our evaluation and Section~\ref{sec:res} evaluates the performance of \algo~using these datasets. The last section concludes the paper.

\section{The Neural Network Architecture}
\label{back}
In deep learning, a recurrent neural network (RNN) layer is a layer type that allows for the processing of sequential data such as a time series by maintaining a memory state that can store information about the recent past. It consists of a single time step of the RNN, which involves computing a hidden state vector $h_t$ and an output vector $y_t$ at each time step $t$. The vector $h_t$ depends not only on the input vector $x_t$ at time step $t$, but also on the hidden state vector $h_{t-1}$ at the previous time step. This dependence allows the network to maintain a memory of past inputs and use this information to inform its current output.

One limitation of this RNN layer is that it can have difficulty remembering long-term dependencies in the input sequence. To overcome this difficulty, the long short-term memory (LSTM) \cite{Hochreiter1997} layer was developed to handle long-term dependencies in the input sequence more effectively.

\subsection{Long-Short-Term Memory (LSTM) Layer}
The LSTM layer consists of the following equations, where $\odot$ represents an element-wise product: 
\begin{align}
    i_t &= \sigma(W_{ii} x_t + b_{ii} + W_{hi} h_{t-1} + b_{hi}) \notag \\
    f_t &= \sigma(W_{if} x_t + b_{if} + W_{hf} h_{t-1} + b_{hf}) \notag \\
    g_t &= \tanh(W_{ig} x_t + b_{ig} + W_{hg} h_{t-1} + b_{hg}) \notag \\
    o_t &= \sigma(W_{io} x_t + b_{io} + W_{ho} h_{t-1} + b_{ho}) \notag \\
    c_t &= f_t \odot c_{t-1} + i_t \odot g_t \notag \\
    h_t &= o_t \odot \tanh(c_t)
    \label{eqn:lstm_h}
\end{align}

Each term in the LSTM equations is described below:
\begin{itemize}
\item $x_t$: The input vector at time $t$.
\item $h_{t-1}$: The previous hidden state vector.
\item $i_t$, $f_t$, $g_t$, $o_t$: The input gate, forget gate, cell gate, and output gate activation vectors, respectively.
\item $c_t$: The memory cell content vector, containing old memory cell content and newly added cell content.
\item $W_{ii}$, $W_{if}$, $W_{ig}$, $W_{io}$: The weight matrices for input gates, forget gates, cell gates, and output gates for the input vector.
\item $W_{hi}$, $W_{hf}$, $W_{hg}$, $W_{ho}$: The weight matrices for the input gates, forget gates, cell gates, and output gates for the previous hidden state.
\item $b_{ii}$, $b_{if}$, $b_{ig}$, $b_{io}$: The bias vectors for the input gates, forget gates, cell gates, and output gates for the input vector
\item $b_{hi}$, $b_{hf}$, $b_{hg}$, $b_{ho}$: The bias vectors for the input gates, forget gates, cell gates, and output gates for the previous hidden state.
\item $h_t$: The hidden state at time $t$.
\end{itemize}

The LSTM network has a cell state that can store information for long periods of time, and three gates that control the flow of information: input gate, forget gate, and output gate. The input gate controls the input to the cell state, the forget gate controls how much of the previous cell state is retained, and the output gate controls the output from the cell state.

At each time step, the LSTM network takes an input $x_t$, the previous hidden state $h_{t-1}$ and the previous cell state $c_{t-1}$, and uses these to compute the input gate $i_t$, forget gate $f_t$, cell gate $g_t$, and output gate $o_t$. 

The cell state $c_t$ is updated based on the input gate $i_t$, forget gate $f_t$, and cell gate $g_t$. The input gate controls how much new information is added to the cell state and the forget gate controls how much old information is retained. The cell gate controls what new information is added to the cell state, by applying an activation function (i.e. $tanh$) to the input and previous hidden state.

Finally, the output gate $o_t$ controls how much of the current cell state is output as the new hidden state $h_t$. The new hidden state is computed by applying the $tanh$ function to the updated cell state $c_t$ and then multiplying it by the output gate $o_t$. The hidden state now contains both short and long-term memory, making it the ideal choice for a unique latent description. 

\subsection{Convolutional Neural Network LSTMs (CNN+LSTMs)}
Convolutional Neural Networks (CNNs) have been successful at image recognition because of their locality bias, which assumes that nearby pixels are useful in identifying an object. The key component of a CNN responsible for this locality bias is the convolutional layer, which convolves a set of filters to the input data in order to extract local features. The filters are typically small in size and slide over the input data in a sequential, linear fashion. This results in a feature map that highlights patterns in the input data and these patterns have the property of translational invariance (i.e. moving a cat a few pixels over still makes the cat present in the image). 

A CNN can also be combined with an LSTM layer by piping the output of the convolutional layer into  the LSTM. We call this hybrid a CNN+LSTM, which is well-suited for discovering patterns in RF transmissions, which have cyclic patterns over time that are predictive of the device.

\section{Methodology}
\label{sec:sys}
\begin{figure}
    \centering
    \includegraphics[width=0.95\linewidth]{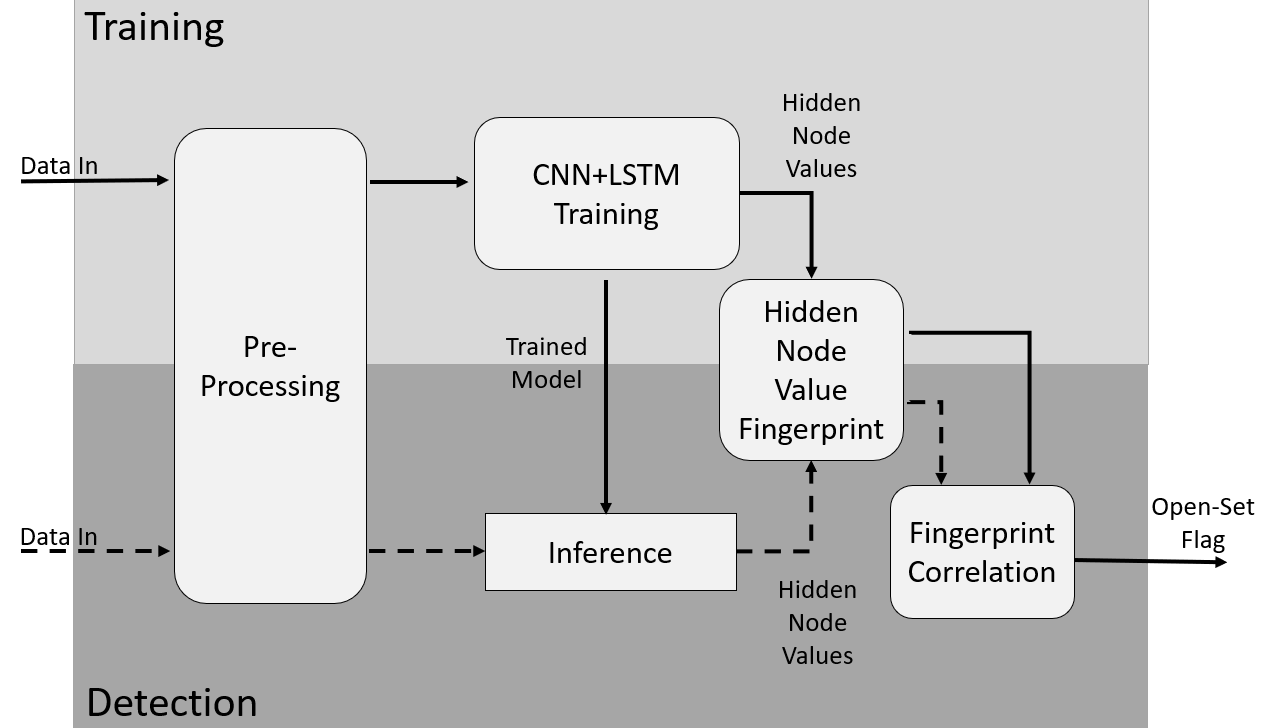}
    \caption{The proposed ML architecture of \algo.}
    \label{fig:system}
\end{figure}

Figure \ref{fig:system} provides an overview of the entire $\algo$ algorithm and illustrates how each component interacts with the others. The top half shows how the training data is processed and the bottom half represents the detection phase operating on test data. 

\subsection{Pre-Processing}\label{subsec:pre}
The data captured from IoT devices during testing is initially processed and stored in the In-phase and Quadrature (IQ) format. The IQ components of an RF signal are crucial in accurately reproducing the original signal and are represented as complex numbers, with the real and imaginary values represented by I and Q, respectively. During testing, each IoT device sends a 20-second message, which is captured by an USRP receiver and saved in a complex number format.

To pre-process the data for analysis, the complex numbers are converted back into their I and Q parts and then segmented into non-overlapping time windows of 2048 samples which we call a \emph{slice}. A signal correlation function is then run on each of the 2048 I and Q samples, each correlated with itself (I to I and Q to Q) to produce the auto-correlation at lags 0 to 2047. The resulting $(2 \times 4096)$ matrix emphasizes cyclostationary features, which are a key part of RF fingerprinting. This new slice contains a mirror image as a result of auto-correlation, so the first half ($2 \times 2048$) is used as the modified feature set (i.e. slice) for training.




\subsection{Training}
The architecture for the CNN+LSTM is shown in Table \ref{tab:architecture}. We train the model with the ADAM optimizer at a fixed learning rate (0.0001) with a cross-entropy loss function. We will discuss hyper-parameter tuning in Section \ref{sec:results.algos}.




\begin{table}[ht]
\centering
\caption{\algo's CNN+LSTM architecture. Notation: Conv2d({\it channels in}:{\it channels out}, {\it kernel dims}), BNorm2D({\it num features}), MaxPool2d({\it pool dims})}
\renewcommand{\arraystretch}{1.3}
\begin{tabular}{|l|}
\hline
\textbf{Layer} \\
\hline \hline
Conv2d (1:16, 2x256) $\shortrightarrow$
BNorm2d(16) $\shortrightarrow$
ReLU $\shortrightarrow$
Dropout(10\%) \\
\hline
Conv2d (16:16, 2x256) $\shortrightarrow$
BNorm2d(16) $\shortrightarrow$
ReLU \\
\hline
Conv2d (16:32, 2x256) $\shortrightarrow$
BNorm2d(32) $\shortrightarrow$
ReLU $\shortrightarrow$
Dropout(10\%) \\
\hline
Conv2d (32:32, 2x256) $\shortrightarrow$
BNorm2d(32) $\shortrightarrow$
ReLU $\shortrightarrow$
MaxPool2d(2x2) \\
\hline
LSTM(64) $\shortrightarrow$
Fully Connected $\shortrightarrow$
LogSoftmax \\
\hline
\end{tabular}

\label{tab:architecture}
\end{table}

\subsection{Detection}
During the detection phase, the IQ data is pre-processed in the same way as in training. Each slice is passed through the trained CNN+LSTM and the final transition in the LSTM layer is extracted. The final transition was determined to be the most suitable for analysis due to the fact that at this point, the LSTM has processed all prior information within the slice. As a result, the internal nodes of the LSTM, specifically the forget gate and cell state, now contain both the long-term and short-term memory associated with the entire slice. This encoding effectively represents the transmission of the device during this specific time slice and is used to create a unique fingerprint. 

\subsection{Hidden State Value Fingerprinting}
\begin{figure}
     \subcaptionbox{Device 5 Fingerprint\label{subfig-1:c5}}{%
       \includegraphics[width=0.233\textwidth, height = 0.237\textwidth]{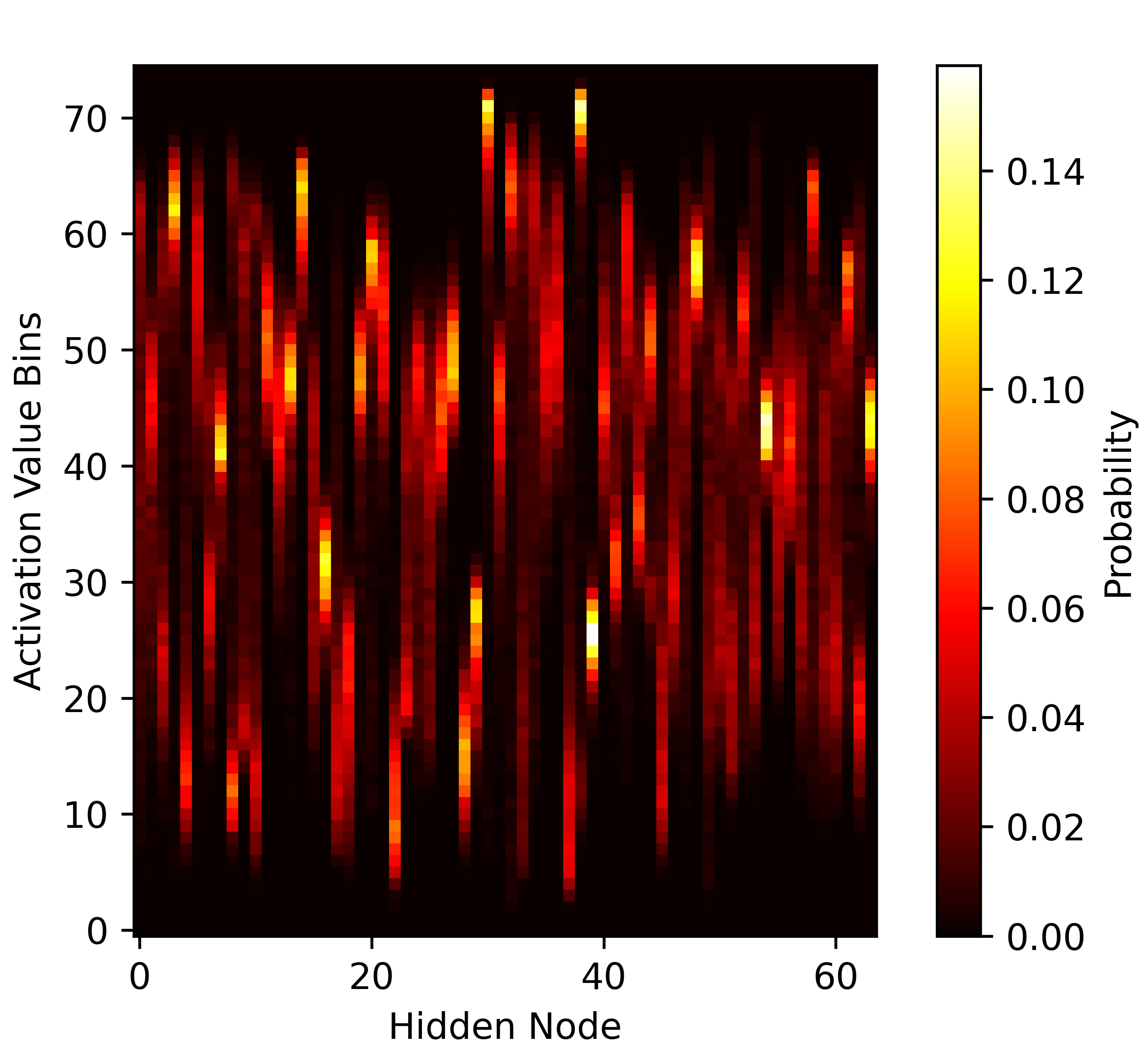}
     }
     \subcaptionbox{Device 6 Fingerprint\label{subfig-2:c6}}{%
       \includegraphics[width=0.233\textwidth, height = 0.237\textwidth]{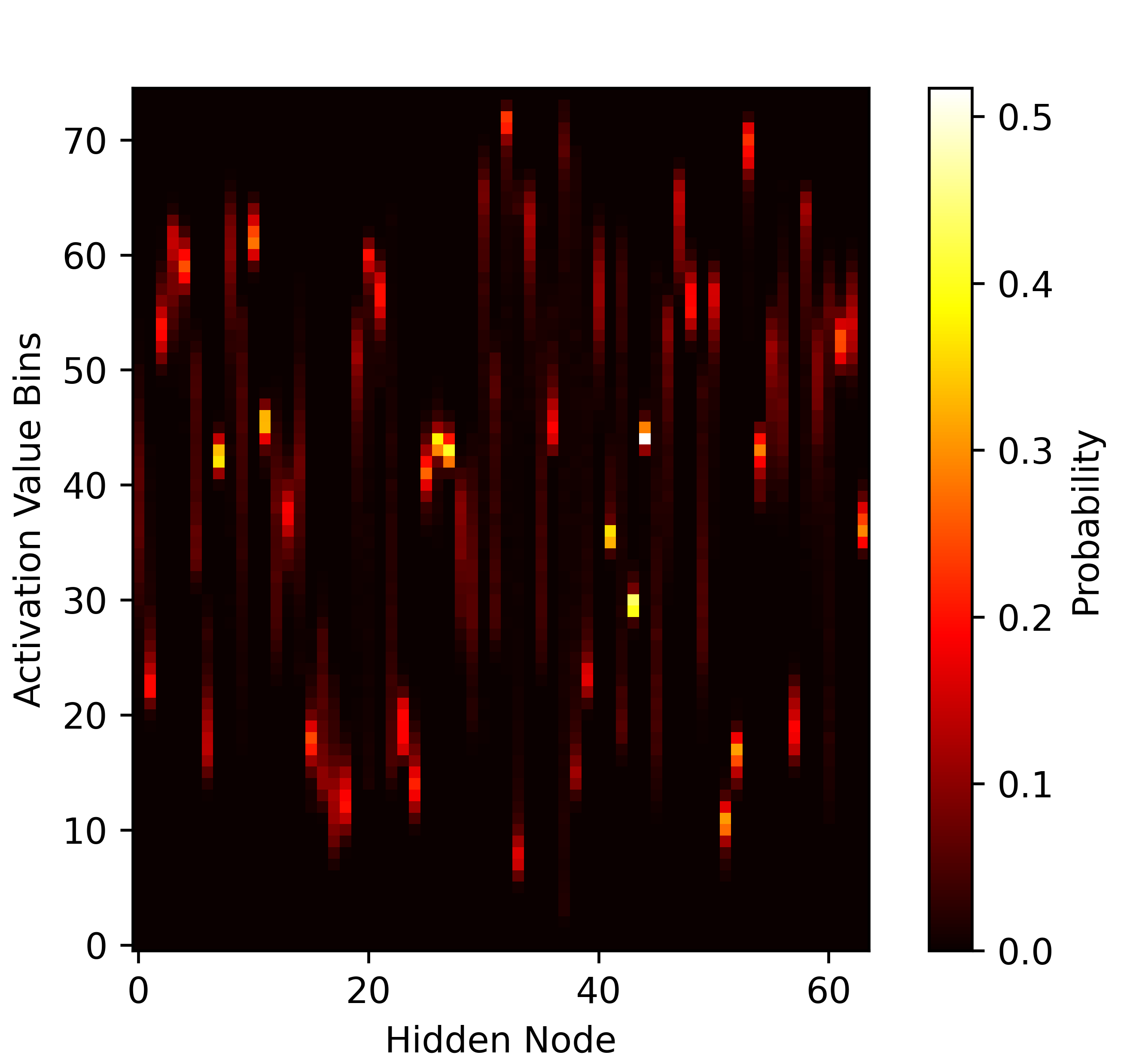}
     }
     \caption{Two unique fingerprints using \algo~under the Wireless-WiFi Dataset (described in Sec.~\ref{sec:dataset}).}
     \label{fig:fingerprints}
\end{figure}

Algorithm \ref{alg:cap} shows how \algo~uses the hidden state values within a trained CNN+LSTM to produce a unique fingerprint for each device in the training set. The first step involves aggregating, for each known device, the hidden node values from all the correctly classified slices during training. Then, for each known device, a histogram with $B$ bins is built that describes the distribution of the hidden state values (i.e. $h_t$ in (\ref{eqn:lstm_h})) for each hidden layer node in the LSTM. With $M$ hidden state nodes, this histogram will be a ($M \times B$) matrix for each device, which serves as the unique fingerprint for that device. Examples of these fingerprints are shown in Fig.~\ref{fig:fingerprints}. 



\begin{algorithm}
\caption{The Fingerprint Generation Algorithm}
\label{alg:cap}
\begin{algorithmic}[1]
\Require $H$ \Comment{Hidden node values from correctly classified training slices}
\State $FP \gets zeroes(\mathit{K_{known} \times M \times B})$ 
\For{$k \gets 0$ \textbf{to} $(K_{known}-1)$} 
\For{$m \gets 0$ \textbf{to} $(M-1)$} \comment{\iterate over hidden nodes}
\State $H_{k,m} \gets H[k,m]$ \Comment{subset of $H$ for device k and node m}
\State $\mathit{FP}[k, m, :] \gets $ Histogram($H_{k,m}$,B)
\EndFor
\EndFor
\State $\textbf{return FP}$
\end{algorithmic}
\hrulefill
\begin{enumerate}
    \item $K_{known}$: the number of closed-set devices
    \item $M$: the number of hidden nodes
    \item $B$: the number of bins in the histogram
    \item $Histogram(Values,B)$: Creates a histogram for $Values$ with $B$ bins 
\end{enumerate}
\end{algorithm}

\subsection{Open-set Fingerprint Correlation}
A number of different approaches can be used to compare test set device fingerprints to the fingerprints of the known devices. For instance, we could compute the probability of a test slice belonging to the histogram for that device, since the histogram is a valid probability distribution. We experimented with different approaches and found that correlations produced the best results. The most common approach for measuring correlation is Pearson's correlation coefficient, which makes a strong assumption that the relationship between two variables is linear. To avoid this strict assumption, we investigated Kendall's $\tau$ \cite{Kendall1938}, which is a non-parametric measure of correlation that quantifies the rank-order association between two variables. 

To compute Kendall's $\tau$, let $\bm{fp_i} = (fp_i^1, \ldots, fp_i^{M*B})$ be the $M*B$ features (i.e. matrix values) for the fingerprint for the $i$th known device. Furthermore, let $\bm{fp}_j = (fp_j^1, \ldots, fp_j^{M*B})$ be the $M*B$ matrix values for the fingerprint of the $j$th device seen in the test set. Kendall's $\tau$ measures the rank correlation in terms of the ranks of the magnitudes of the features $(fp_i^1, \ldots, fp_i^{M*B})$ and $(fp_j^1, \ldots, fp_j^{M*B})$. Specifically, two feature indices $i1$ and $i2$ are said to be \emph{concordant} if $fp_i^{i1} > fp_i^{i2}$ and $fp_j^{i1} > fp_j^{i2}$ (or equivalently if $fp_i^{i1} < fp_i^{i2}$ and $fp_j^{i1} < fp_j^{i2}$), otherwise they are said to be \emph{discordant}. Computing Kendall's $\tau$ (see (\ref{eqn:tau})) requires the number of concordant ($P$) and discordant pairs ($Q$), as well as the number of tied pairs of feature indices only in $\bm{fp_i}$ ($T$) and only in $\bm{fp_j}$ ($U$). 
\begin{equation}
\tau = \frac{P - Q}{\sqrt{(P + Q + T) \cdot (P + Q + U)}}
\label{eqn:tau}
\end{equation}
We chose Kendall's $\tau$ because it produced significantly better performance than a linear correlation.

Algorithm~\ref{alg:openset-detector} illustrates the unknown device detection process. Each test device has its slices converted to a test fingerprint, which is an $M \times B$ histogram. The test fingerprint for the $k$th test device was compared to all the known fingerprints, and its maximal rank correlation coefficient $\tau_k^{*}$ was computed. We use $(1-\tau_k^{*})$ to indicate the degree to which the test device was not correlated to a known device. If the value $(1 - \tau_k^*)$ was above a threshold, an open-set flag was raised. 

\begin{algorithm}
\caption{The Open-Set Detector}
\label{alg:openset-detector}
\begin{algorithmic}[1]
\Require $FP$ \Comment{Fingerprint Tensor from Algo.~\ref{alg:cap}}
\Require $H_{test} \gets \mathbf{K_{test}\times M \times S_{test}}$
\Require $FP_{test} \gets zeroes(\mathbf{K_{test} \times M \times B})$ 
\Require $result \gets zeroes(\mathbf{K_{test}})$
\For{$k \gets 0 \textbf{ to } (K_{test}-1)$}
\For{$m \gets 0 \textbf{ to } M-1$}
\State $FP_{test}[k,m,:] \gets Histogram(H_{test}[k,m,:],B)$ 
\EndFor
\EndFor

\For{$k \gets 0 \textbf{ to } (K_{test}-1)$}
    \For{$l \gets 0 \textbf{ to } (K_{known}-1) $}
    \State $\tau_{k,l} = KT(flatten(FP[l]), flatten(FP_{test}[k]))$
    \EndFor 
    \State $\tau_{k}^{*} = \underset{l}{\mathit{max}} (\tau_{k,l})$
    \State $\mathit{result[k]} = (1 - \tau_{k}^{*})$
\EndFor
\State $\textbf{return result}$
\end{algorithmic}
\hrulefill
\begin{enumerate}
    \item $K_{test}$: the total number of test devices
    \item $M$: the number of hidden nodes
    \item $S_{test}$: the number of test slices per device
    \item $B$: the number of bins in the histogram
    \item $H_{test}$: the hidden state values for the test slices
    \item $FP_{test}$ : the test fingerprints
    \item $K_{known}$ : the number of known devices
    \item KT: Kendall Tau correlation function
    \item $flatten$: function to flatten 2D matrix to 1D vector
    \item $\tau_k$: The rank correlation coefficient for device $k$
    \item $result$: the per-device vector of unthresholded predictions (higher is more indicative of an unknown device)
\end{enumerate}
\end{algorithm}



\section{Testbed and Datasets}
\label{sec:dataset}
\begin{figure}
     \subfloat[15 Pycom Transmitters.\label{subfig-1:tx}]{%
       \includegraphics[width=0.2\textwidth, height = 0.237\textwidth]{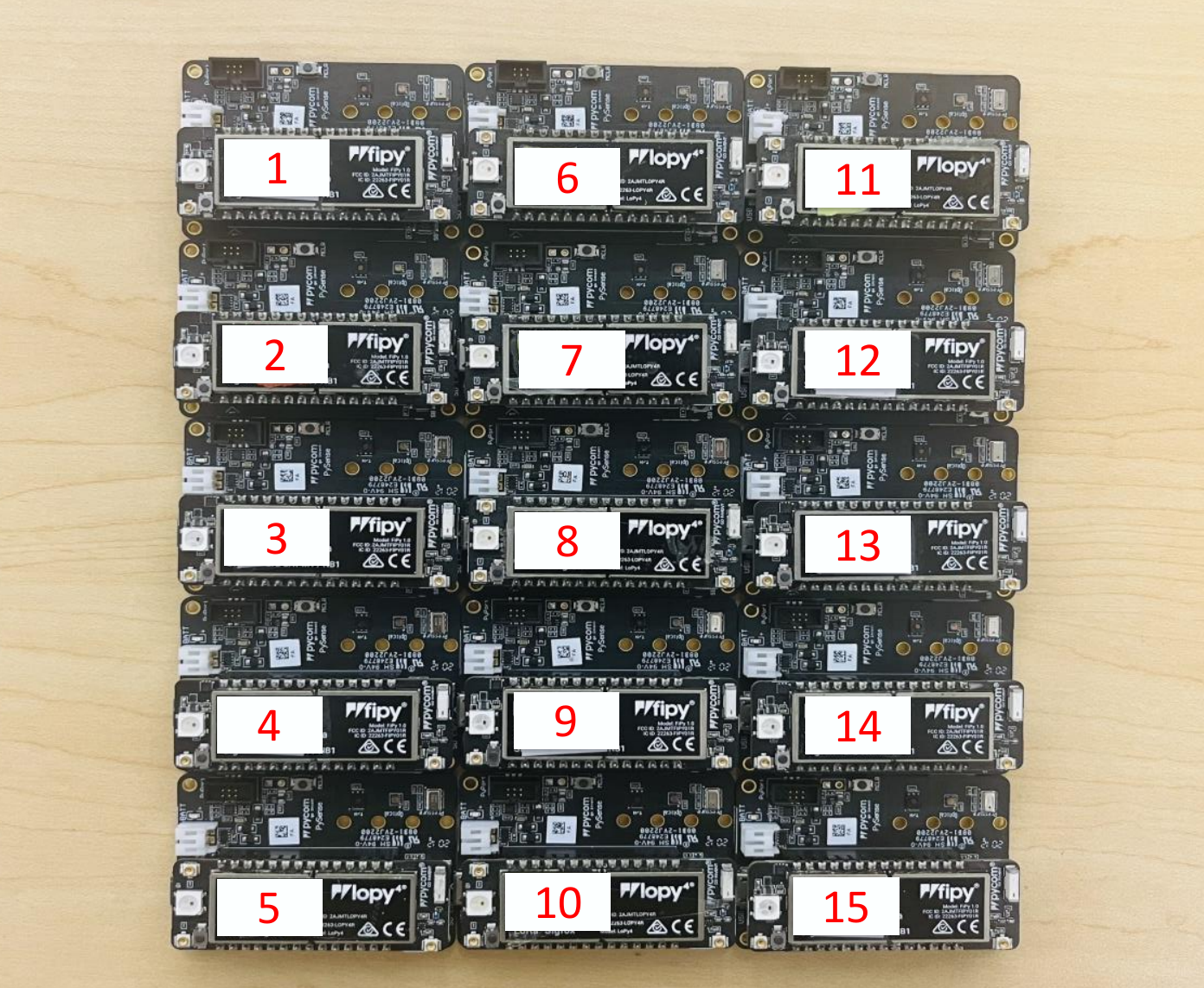}
     }
    \hspace{0.00001cm}
     \subfloat[{Wired-WiFi Vs. Wireless-WiFi.}\label{subfig-2:recv}]{%
       \includegraphics[width=0.25\textwidth, height = 0.237\textwidth]{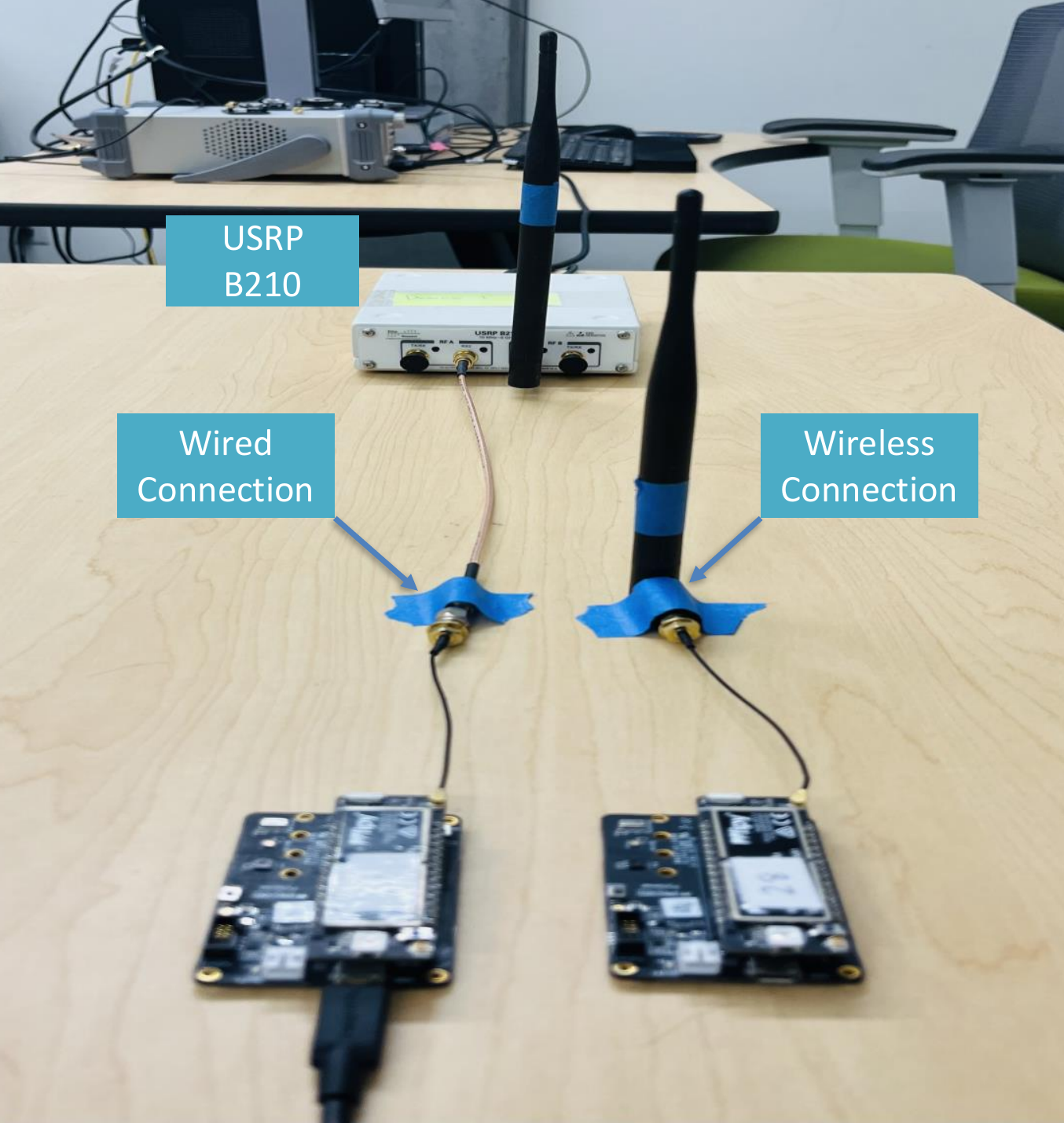}
     }
     \caption{IoT Testbed consisting of 15 Pycom transmitting devices and a USRP B210 receiving device.}
     \label{fig:testbed}
\end{figure}

In this work, we utilized three RF datasets: LoRa, Wireless-WiFi, and Wired-WiFi which have been collected using a testbed of $15$ PyCom IoT devices as transmitters: $9$ Fipy boards and $6$ Lopy$4$ boards on top of PySense sensor shields (pictured in Fig.~\ref{subfig-1:tx}). On the reception side, we used an Ettus USRP (Universal Software Radio Peripheral) B$210$ with a VERT$900$ antenna for the data acquisition. For the LoRa dataset, we captured the LoRa transmissions of a duration of $20$s each, in an indoor environment where the devices were located $5$m away from the receiver. Each Pycom device was connected to a dedicated LoRa antenna and configured to transmit LoRa transmissions at the $915$MHz and $125$KHz bandwidth. These transmissions have been sampled by the USRP receiver at a rate of $1$MSps. Refer to the Indoor LoRa dataset section in \cite{deployment_variability,elmaghba_LoRa} for more details. 

For the WiFi datasets, the same Pycom devices were programmed to transmit WiFi IEEE802.11B frames at a center frequency of $2.412$GHz and $20$MHz bandwidth. These frames have been sampled and digitally down-converted by the same USRP receiver at a sample rate of $45$MSps. Each WiFi capture lasts for $2$ minutes generating more than $5000$ frames per device where each frame consists of $25170$ complex-valued samples. While the transmitters were located $1$m away from the receiver and connected to the same antenna in the wireless WiFi dataset, a $12$inch SMA cable was used to connect them directly to the USRP receiver in the wired WiFi dataset as shown in Fig.~\ref{subfig-2:recv}.

\section{Results and Discussion}
\label{sec:res}
For each of the three studied datasets, we set up 3 experiments in which we randomly selected 10 devices to be the known devices and 5 devices to be the unknown devices. We then evaluate our approach using a variant of 5-fold cross-validation designed to handle evaluation of open-set detection.
We use a dataset with an equal number of data samples (i.e. slices) from each of the 15 devices. 
We divide each device's data into 5 non-overlapping equally-sized partitions. Under the traditional cross-validation process, in each fold of cross-validation, 4 of the partitions for that device are used as the training set while the remaining partition is used as the test set. The partitions are reassigned to training and testing in the other folds, such that each fold ends up using a different partition for testing, with no overlap between test sets for each fold. Data from the 10 known devices follow this traditional 5-fold cross-validation process. The main difference in our variant occurs with the test partition in each fold. In open-set detection, the test set contains both the test partition for the 10 known devices as well as the test partition for the 5 unknown devices. We emphasize that in each fold, the data from the 5 unknown devices are only seen during testing and never seen during training.

Thus, to summarize the overall process, in each fold of cross-validation, \algo~is trained on the training set. After training, we generated 10 device fingerprints using the correctly classified samples from the 4 partitions of the known device training data. During the detection phase, \algo~takes each test sample from the test partition and compares it to the 10 known device fingerprints to perform a binary prediction as to whether or not the sample belongs to a known or unknown device.



\begin{table*}[t]
\caption{Average Test AUPRC for \algo~vs. other algorithms on the (a) LoRa, (b) Wireless-WiFi and (c) Wired-WiFi datasets. The graphs on top plot the results in the tables below. Statistical significance is indicated with * in the tables.}

\begin{subtable}{0.66\columnwidth}
\centering
\renewcommand{\arraystretch}{1.2}
\includegraphics[height=.15\textheight]{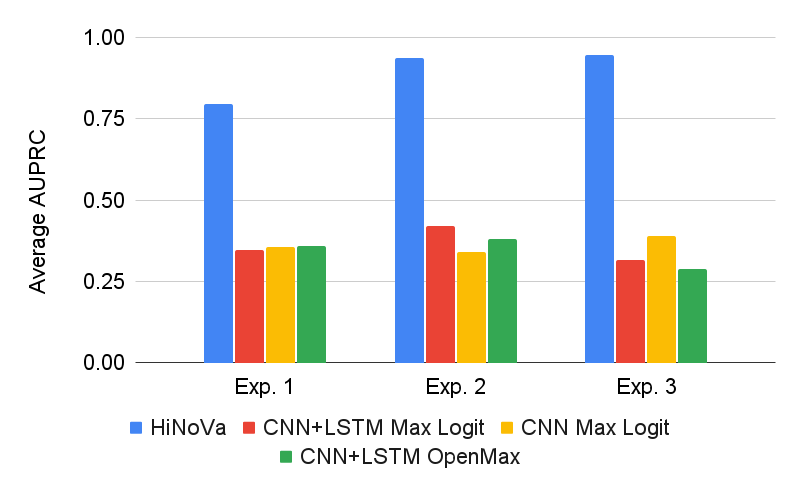}  
\begin{tabular}[t]{|p{1.5cm}||c|c|c|}
\hline
& \textbf{Exp. 1} & \textbf{Exp. 2} & \textbf{Exp. 3} \\ \hline
\textbf{\algo} & \textbf{0.80*} & \textbf{0.94*} & \textbf{0.95*} \\ \hline
\textbf{CNN+LSTM MaxLogit} & 0.35 & 0.42 & 0.32 \\ \hline
\textbf{CNN MaxLogit} & 0.36 & 0.34 & 0.39 \\ \hline
\textbf{CNN+LSTM OpenMax} & 0.36 & 0.38 & 0.29 \\ \hline
\end{tabular}
\vspace{0.1pt}
\caption{LoRa}
\label{tab:auprc_lora}
\end{subtable}
\hfill
\begin{subtable}{0.66\columnwidth}
\centering
\renewcommand{\arraystretch}{1.2}
\includegraphics[height=.15\textheight]{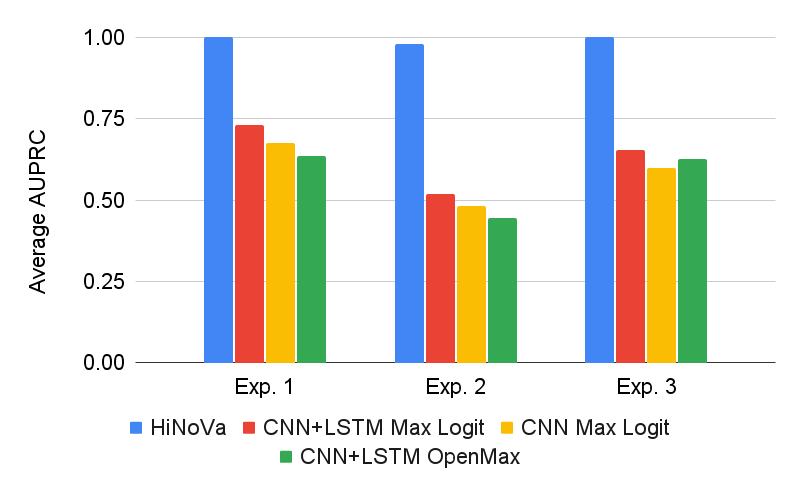}
\begin{tabular}[t]{|p{1.5cm}||c|c|c|}
\hline
& \textbf{Exp. 1} & \textbf{Exp. 2} & \textbf{Exp. 3} \\ \hline
\textbf{\algo~} & \textbf{1.00*} & \textbf{0.98*} & \textbf{1.00*} \\ \hline
\textbf{CNN+LSTM MaxLogit} & 0.73 & 0.52 & 0.65 \\ \hline
\textbf{CNN MaxLogit} & 0.68 & 0.48 & 0.60 \\ \hline
\textbf{CNN+LSTM OpenMax} & 0.63 & 0.45 & 0.63 \\ \hline
\end{tabular}
\vspace{0.1pt}
\caption{Wireless-WiFi}
\label{tab:auprc_wifi}
\end{subtable}
\hfill
\begin{subtable}{0.66\columnwidth}
\centering
\renewcommand{\arraystretch}{1.2}
\includegraphics[height=.15\textheight]{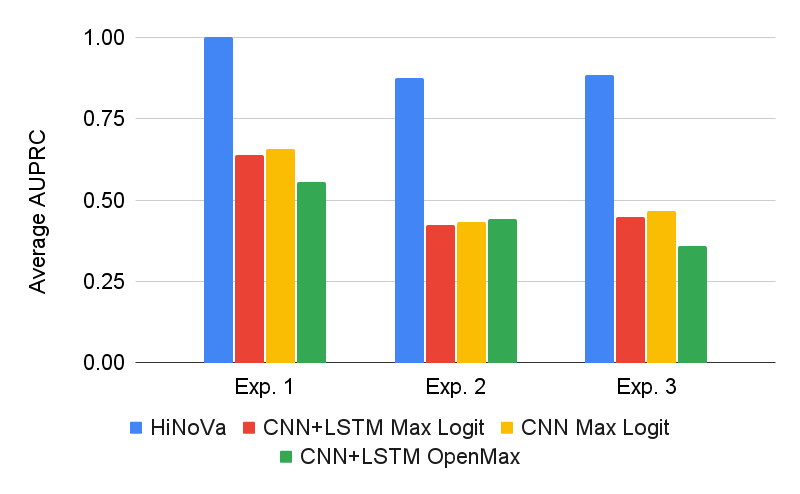}
\begin{tabular}[t]{|p{1.5cm}||c|c|c|} 
\hline
& \textbf{Exp. 1} & \textbf{Exp. 2} & \textbf{Exp. 3} \\ \hline
\textbf{\algo~} & \textbf{1.00*} & \textbf{0.87*} & \textbf{0.89*} \\ \hline
\textbf{CNN+LSTM MaxLogit} & 0.59 & 0.43 & 0.45 \\ \hline
\textbf{CNN MaxLogit} & 0.63 & 0.43 & 0.47 \\ \hline
\textbf{CNN+LSTM OpenMax} & 0.56 & 0.44 & 0.36 \\ \hline
\end{tabular}
\vspace{0.1pt}
\caption{Wired-WiFi}
\label{tab:auprc_wired}
\end{subtable}%
\end{table*}

\subsection{Algorithms and Performance Metrics}
\label{sec:results.algos}
We compare \algo~against a number of other open-set detection methods:

\noindent {\bf 1) CNN model using MaxLogit}: This baseline uses a CNN augmented with the MaxLogit process for detecting open set instances. As was pointed out in a recent work \cite{vaze22}, MaxLogit, though simple, is a strong open-set detector.

\noindent {\bf 2) CNN+LSTM model using MaxLogit}: The previous baseline interprets each observation in a slice as an i.i.d. data instance. In reality, the observations in a slice have a sequential relationship and using a CNN+LSTM instead of a CNN enables the detector to model these sequential relationships. As before, we use the MaxLogit approach for open-set detection.

\noindent {\bf 3) OpenMax \cite{Bendale2015}}: This common baseline reweights the activation vectors that are input to the final Softmax layer of the NN to better distinguish between known and unknown devices. The weighting function is based on a Weibull distribution, which is used to model extreme values and is used in OpenMax to model the right tail of the activation distribution corresponding to the highest activation values. OpenMax only reweights the activations for the top $\alpha$ classes with the highest activation values.

\noindent {\bf 4) Akar \cite{akar_open_2022}}: The work by Akar et al. \cite{akar_open_2022} is a state-of-the-art open-set detector specifically for time series. We refer to this approach as \emph{Akar}. The Akar method uses Dynamic Time Warping (DTW) to compute the similarity between a test set time series and the barycenters of known devices. 



We use AUPRC (Area Under Precision-Recall Curve) as the evaluation metric~\cite{saito2015precision}. AUPRC is a suitable metric for the open-set detection problem because there can be a significant data imbalance between known and unknown devices. In our setup, we have twice as much data from known devices than from unknown devices during testing. AUPRC considers the trade-off between precision and recall across a range of detection thresholds and yields an overall threshold-independent summary statistic of the detector's performance. 

To determine the hyper-parameter settings for our deep learning models, we use post-hoc tuning on CNN+LSTM MaxLogit.  We use CNN+LSTM MaxLogit because parts of its architecture are shared with CNN MaxLogit and CNN+LSTM OpenMax. Post-hoc tuning refers to looking at the performance of CNN+LSTM MaxLogit on the test set; this is technically giving CNN+LSTM MaxLogit an unfair advantage as it is allowed to see the test set, but we will show that even with this advantage, \algo~still outperforms the MaxLogit models by a significant margin.

Specifically, we post-hoc tune the kernel size ($2 \times 256$) and dropout rate (10\%) in the CNN layer to achieve high accuracy in closed set classification using a grid search. Attaining good closed set accuracy has recently been shown to produce good open set detectors \cite{vaze22}. We also post-hoc tune the number of hidden nodes to achieve high AUPRC for the open-set prediction task for CNN+LSTM MaxLogit. The resulting values of these hyperparameters were applied to \algo, which clearly puts it at a disadvantage because these hyperparameters were tuned for a completely different algorithm (i.e. CNN+LSTM MaxLogit), but \algo~still performs well.

We evaluated \algo~with 25, 50, 75 and 100 bins and found that it resulted in small differences in AUPRC $(< 0.03)$. We report results with 25 bins in our experiments.

\subsection{Experimental Results}
Our performance evaluation is done using three different RF datasets: LoRa, Wireless-WiFi, and Wired-WiFi, as described in Sec.~\ref{sec:dataset}.  
Tables~\ref{tab:auprc_lora}, \ref{tab:auprc_wifi} and \ref{tab:auprc_wired} show the average AUPRC values for the LoRa, Wireless-WiFi, and Wired-WiFi datasets respectively. \algo~consistently outperformed the other methods, achieving statistically significant results (Wilcoxon Signed Rank Test, $\alpha = 0.05$) in all three experiments.
CNN+LSTM MaxLogit, CNN MaxLogit, and OpenMax lagged behind both \algo~by a substantial gap in AUPRC, with no consistent top performer in this second tier of algorithms. Due to the extensive computational time of Akar, the algorithm did not complete within 24 hrs, making it infeasible to be used for this security use case.



\begin{figure*}
\centering
  \subcaptionbox{LoRa dataset}{
  \includegraphics[width=.66\columnwidth,height=.15\textheight]{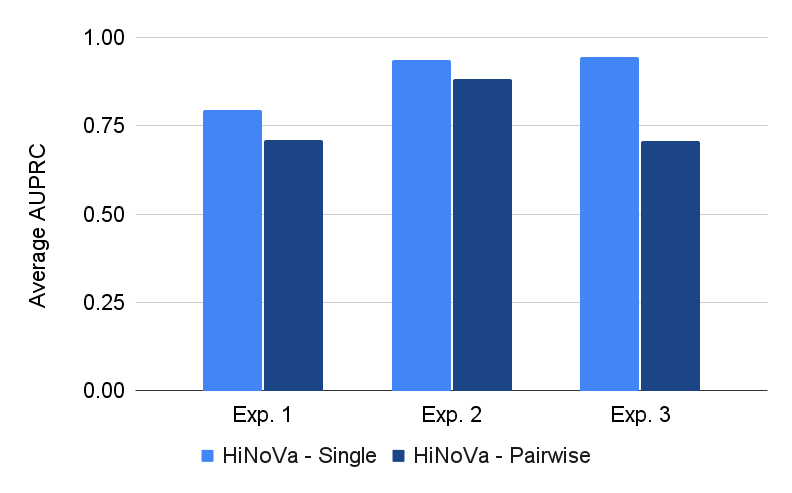}
  }
%
  \subcaptionbox{Wireless-WiFi dataset}{
  \includegraphics[width=.66\columnwidth,height=.15\textheight]{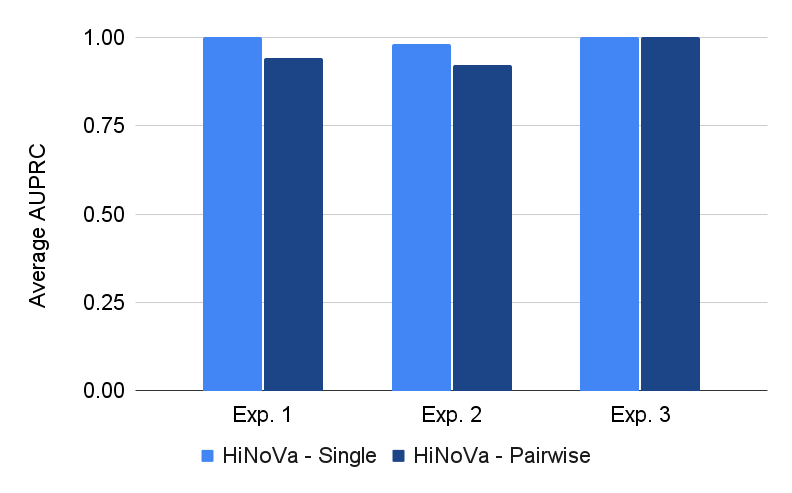}
  }
%
  \subcaptionbox{Wired-WiFi dataset}{
  \includegraphics[width=.66\columnwidth,height=.15\textheight]{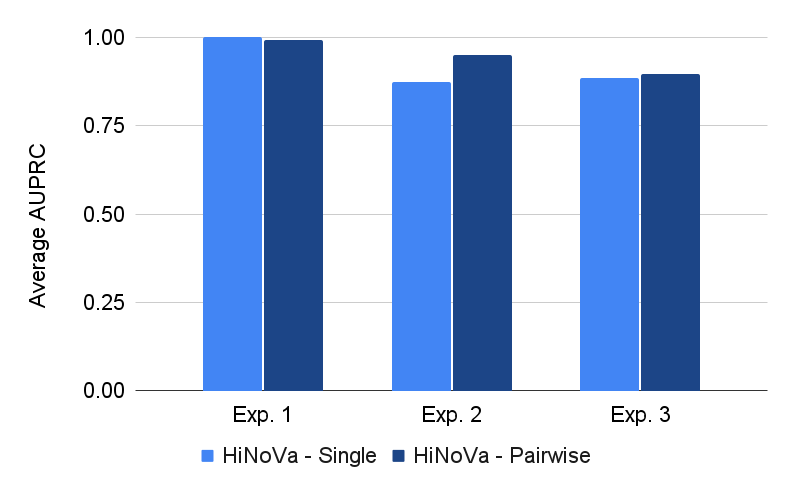}
  }

\caption{Average test AUPRCs for the single hidden node detector vs. the pairwise hidden node detector.}

\label{fig:auprc_wired_transitions}

\end{figure*}

Overall, the results suggest that \algo~is an effective detector of unknown devices using LoRa, Wireless-WiFi and Wired-Wifi protocols, outperforming other methods by a significant margin. The hidden state values correspond to a compact representation of the autocorrelation lags in the IQ data within a slice, and the distribution of this representation, as represented in the histogram used to derive the fingerprint, provides an effective summary of the device-specific information that \algo~is able to leverage. Finally, the MaxLogit approaches and the OpenMax approach only rely on the logits of the penultimate layer of the NN. These logits, which are used to derive the output probabilities from the NN, lack the information contained in the fingerprints and are thus less effective at identifying unknown devices.

\subsection{Pairwise vs Single Hidden Node Values}
Since LSTMs use the hidden node value from the previous time step ($h_{t-1}$) to compute the value of the current hidden node ($h_t$), we explore building the RF fingerprint with the pair of hidden node values at consecutive times $(h_{t-1}, h_t)$ instead of the hidden node value at a single time ($h_t$). Figure \ref{fig:auprc_wired_transitions} compares the performance of a single vs pairwise hidden node value detector. Figure \ref{fig:auprc_wired_transitions} shows that for \algo, the results are mixed, with a pairwise detector outperforming the single node detector in about half of the experiments. These results indicate that pairwise transitions can have predictive value in some cases, but in other cases, this transition is simply noise. Given the additional computational cost of the pairwise node detector in both time and memory, we recommend using the single node detector.

\comment{
\begin{figure}
  \centering
  \includegraphics[width=.9\linewidth]{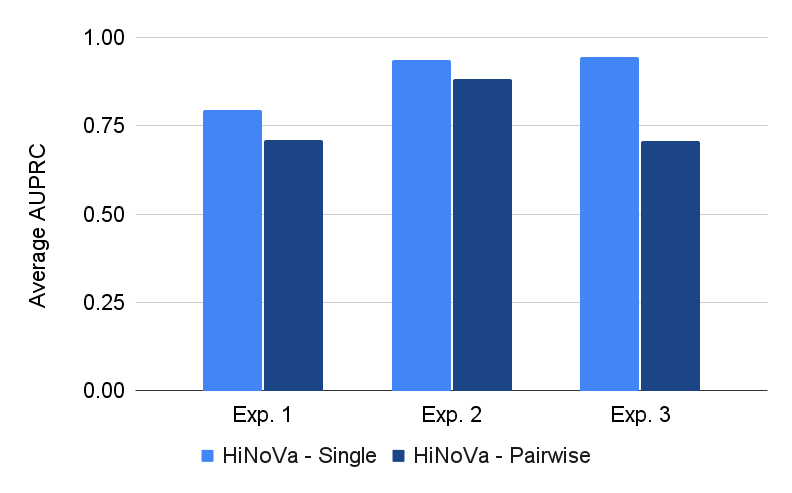}
  \caption{Single Time vs. Pairwise Time Hidden State Activation - LoRa Dataset}
  \label{fig:auprc_lora_transitions}
\end{figure}

\begin{figure}
  \centering
  \includegraphics[width=.9\linewidth]{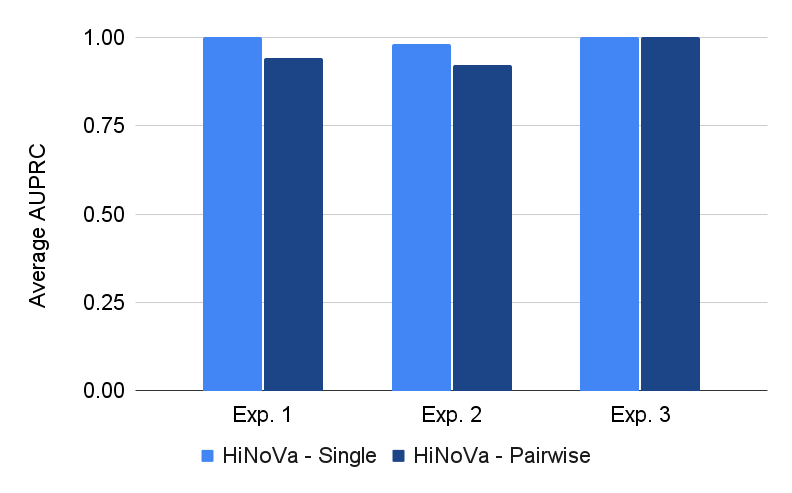}
  \caption{Single Time vs. Pairwise Time Hidden State Activation - WiFi Dataset}
  \label{fig:auprc_wifi_transitions}
\end{figure}

\begin{figure}
  \centering
  \includegraphics[width=.9\linewidth]{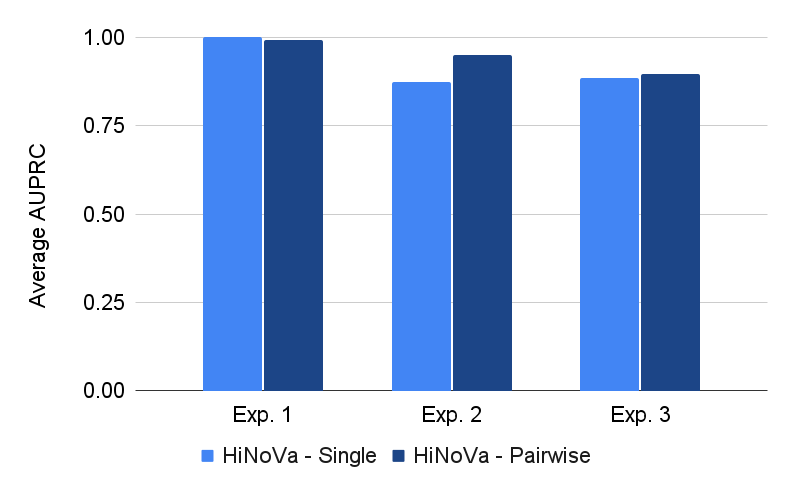}
  \caption{Single Time vs. Pairwise Time Hidden State Activation - Wired Dataset}
  \label{fig:auprc_wired_transitions}
\end{figure}
}

\comment{
\begin{figure*}
  \centering

\begin{subfigure}[h]{width=.5\columnwidth}
  \includegraphics[width=.5\columnwidth]{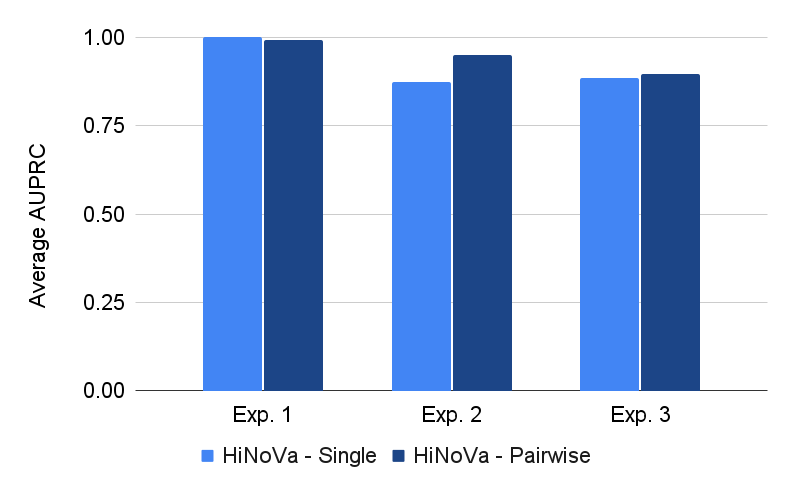}
  \caption{a}
\end{subfigure}

\begin{subfigure}[h]{width=.5\columnwidth}
  \includegraphics[width=.5\columnwidth]{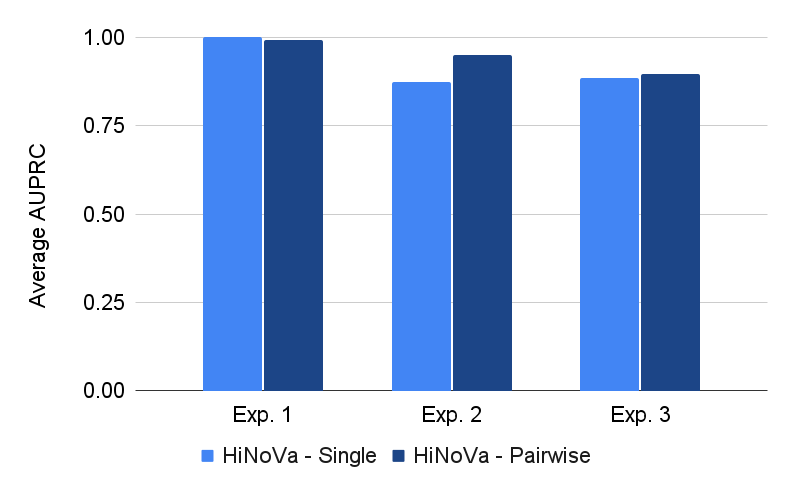}
  \caption{b}
\end{subfigure}

\caption{fig:auprc_wired_transitions}
\end{figure*}
}

\section{Conclusion}
\label{sec:conc}
This work proposed \algo, a novel open-set detection method based on the activation patterns of the hidden states within a CNN+LSTM model. This approach significantly improves the AUPRC on LoRa, Wireless-WiFi, and Wired-WiFi datasets over other open-set detection methods. Additionally, because of its structure, the proposed method can run on standard consumer hardware with minimal setup data and training time. Future work will investigate using attention-based deep learning models. 

\section{Acknowledgements}
\label{sec:ack}
{\small This work is supported in part by Intel/NSF Award No. 2003273.}

\bibliographystyle{IEEEtran}
\bibliography{Thesis}




\end{document}